\documentclass{JHEP3}
\usepackage{epsfig}
\usepackage{epstopdf}
\input{epsf.sty}
\usepackage{epsf}
\usepackage{amssymb}

\renewcommand\({\left(}
\renewcommand\){\right)}
\renewcommand\[{\left[}
\renewcommand\]{\right]}

\newcommand{\ra}{\rightarrow}

\def\lsim{\raise 0.4ex\hbox{$<$}\kern -0.8em\lower 0.62
ex\hbox{$\sim$}}

\def\gsim{\raise 0.4ex\hbox{$>$}\kern -0.7em\lower 0.62
ex\hbox{$\sim$}}

\def\lbar{{\hbox{$\lambda$}\kern -0.7em\raise 0.6ex
\hbox{$-$}}}

\newcommand\eq[1]{eq.~(\ref{#1})}
\newcommand\eqs[2]{eqs.~(\ref{#1}) and (\ref{#2})}

\newcommand\Eqs[2]{Equations~(\ref{#1}) and (\ref{#2})}

\newcommand\eqst[2]{eqs.~(\ref{#1})--(\ref{#2})}
\newcommand\pa{\partial}
\newcommand\p{\partial}

\newcommand\ee{\end{equation}}
\newcommand\be{\begin{equation}}
\def\bea{\begin{array}}
\def\eea{\end{array}}\def\ea{\end{array}}
\newcommand\ees{\end{eqnarray}}
\newcommand\bees{\begin{eqnarray}}

\def\p1{{\bf p}_1}
\def\p2{{\bf p}_2}
\def\k1{{\bf k}_1}
\def\k2{{\bf k}_2}

\newcommand{\dddM}{\kern 0.2em \raise 1.9ex\hbox{$...$}\kern -1.0em \hbox{$M$}}
\newcommand{\dddQ}{\kern 0.2em \raise 1.9ex\hbox{$...$}\kern -1.0em \hbox{$Q$}}
\newcommand{\dddI}{\kern 0.2em \raise 1.9ex\hbox{$...$}\kern -1.0em\hbox{$I$}}
\newcommand{\dddJ}{\kern 0.2em \raise 1.9ex\hbox{$...$}\kern-1.0em
\hbox{$J$}}
\newcommand{\dddcalJ}{\kern 0.2em \raise 1.9ex\hbox{$...$}\kern-1.0em
\hbox{${\cal J}$}}

\newcommand{\dddO}{\kern 0.2em \raise 1.9ex\hbox{$...$}\kern -1.0em
\hbox{${\cal O}$}}
\def\dddz{\raise 1.5ex\hbox{$...$}\kern -0.8em \hbox{$z$}}
\def\dddd{\raise 1.8ex\hbox{$...$}\kern -0.8em \hbox{$d$}}
\def\dddbd{\raise 1.8ex\hbox{$...$}\kern -0.8em \hbox{${\bf d}$}}
\def\ddbd{\raise 1.8ex\hbox{$..$}\kern -0.8em \hbox{${\bf d}$}}
\def\dddx{\raise 1.6ex\hbox{$...$}\kern -0.8em \hbox{$x$}}

\def\D{\Delta}
\def\p{\partial}

\def\nn{\nonumber}

\def\s{\sigma}

\def\G{\Gamma}
\def\d{\delta}

\def\eps{\epsilon}

\def\dslash{\hspace{-1mm}\not{\hbox{\kern-2pt $\partial$}}}
\def\Dslash{\not{\hbox{\kern-4pt $D$}}}
\def\pslash{\not{\hbox{\kern-2.1pt $p$}}}
\def\kslash{\not{\hbox{\kern-2.3pt $k$}}}
\def\qslash{\not{\hbox{\kern-2.3pt $q$}}}

\newcommand{\inT}{\int_{-\infty}^{\infty}}

\newcommand{\Dl}{\int{\cal D}\lambda}

\title{The Halo Mass Function from  Excursion 
Set Theory
with a	Non-Gaussian Trispectrum}
\author{ Michele Maggiore$^1$ and  Antonio Riotto$^{2,3}$\\
$^1$ D\'epartement de Physique Th\'eorique, 
Universit\'e de Gen\`eve,\\ 24 quai Ansermet, CH-1211 Gen\`eve, Switzerland
\vskip 0.3cm 
$^2$ CERN, PH-TH Division, CH-1211, Gen\`eve 23,  Switzerland
\vskip 0.3cm
$^3$INFN, Sezione di Padova, Via Marzolo 8,
I-35131 Padua, Italy
\vskip 0.3cm
Emails: michele.maggiore@unige.ch, antonio.riotto@cern.ch
\vskip 0.3cm
CERN-PH-TH/2009-204}
\abstract{ 
A sizeable level of non-Gaussianity in the primordial cosmological
perturbations may be 
induced by a large trispectrum, i.e.  by a large connected four-point
correlation function.	 
We compute  the effect of a  primordial  non-Gaussian trispectrum on the 
halo mass function,  within excursion set theory. We use the formalism that we
have developed in a previous series of papers and which allows us to take into
account the fact that,
in the presence of non-Gaussianity,  the stochastic evolution of the smoothed
density field, as a function of  the smoothing scale,  is
non-markovian. In the large mass limit, the leading-order term that we find
agrees with the leading-order term of the results found in the literature using
a more heuristic Press-Schecther (PS)-type approach. Our approach however also
allows us to evaluate consistently the  subleading terms, 
which depend not only on
the four-point cumulant but also on derivatives of the four-point correlator, and which cannot be
obtained within non-Gaussian extensions of PS theory. We perform explicitly
the computation up to next-to-leading order.
}

\begin{document}

\maketitle

\section{Introduction}
Over the last
decade a great deal of	evidence has been accumulated from the Cosmic Microwave
Background (CMB) anisotropy and Large Scale Structure (LSS) spectra
that the observed
structures originated from seed fluctuations generated during a primordial
stage of inflation. While standard single-field models of
slow-roll inflation
predict that these fluctuations are very close to  
gaussian (see \cite{acquaviva,maldacena}), 
non-standard scenarios allow for a larger level of non-Gaussianity (NG)
(see \cite{bartoloreview} 
and references therein).  A signal is  gaussian if the information it carries
is completely
encoded in the two-point correlation function, all higher connected correlators
being
zero.
Deviations from Gaussianity are therefore  encoded, e.g., in the connected 
 three- and four-point correlation functions which are dubbed the bispectrum
and the trispectrum,
respectively. A phenomenological way of parametrizing the level of NG is to
expand the fully 
 non-linear primordial Bardeen
 gravitational potential $\Phi$ in powers of the linear gravitational potential
$\Phi_{\rm L}$
 
 \be
 \label{phi}
 \Phi=\Phi_{\rm L}+f_{\rm NL}\left(\Phi_{\rm L}^2-\langle\Phi_{\rm
L}^2\rangle\right)+g_{\rm NL}
 \Phi_{\rm NL}^3\, .
 \ee
The  dimensionless
quantities  $f_{\rm NL}$ and $g_{\rm NL}$  set the
magnitude of the three- and
four-point 
correlation functions, respectively  \cite{bartoloreview}. If the process generating the primordial
non-Gaussianity  
is local in space, the parameter  $f_{\rm NL}$ and $g_{\rm NL}$ in Fourier
space are 
independent of the momenta
entering the corresponding  correlation functions; if instead the process which
generates the
primordial cosmological perturbations is
non-local in space, like in 
models of inflation with non-canonical kinetic terms, $f_{\rm NL}$ and $g_{\rm
NL}$ 
acquire a dependence on the momenta. 
It is clear that  detecting a
significant amount of 
non-Gaussianity and its shape either from the CMB or from the
LSS offers the possibility of opening a   window  into the 
dynamics of the universe
during the very first 
stages of its
evolution. Non-Gaussianities are particularly relevant in the  high-mass end of
the power spectrum of perturbations, i.e. on the scale of galaxy clusters,
since the effect of NG fluctuations becomes especially visible on
the  tail of the probability distribution. 
As a result, both 
the abundance and  the clustering properties of very massive halos
are sensitive probes of primordial 
non-Gaussianities \cite{MLB,GW,LMV,MMLM,KOYAMA,MVJ,RB,RGS}, 
and could be detected or significantly constrained by
the various planned large-scale galaxy surveys,
both ground based (such as DES, PanSTARRS and LSST) and on satellite
(such as EUCLID and ADEPT) see,  e.g.  \cite{Dalal} and \cite{CVM}. 
Furthermore, the primordial non-Gaussianity
alters the clustering of dark matter halos inducing a scale-dependent
bias on large 
scales \cite{Dalal,MV,slosar,tolley}  while even for small primordial
non-Gaussianity the evolution of perturbations on super-Hubble scales yields
extra
contributions on 
smaller scales \cite{bartolosig,MV2009}.  
The strongest current
limits on the strength of local non-Gaussianity set the $f_{\rm NL}$
parameter to be in the range $-4<f_{\rm NL}<80$ at 95\% confidence level
\cite{zal}. 

While  the literature on NG has vastly focussed on the impact on observables
induced by
a non-vanishing bispectrum, only recently attention has been devoted to the 
impact of a non-vanishing trispectrum of cosmological perturbations
\cite{t1,t2,t3,t4,coles}. 
This has been computed in several models, such as multifield slow-roll
inflation model
\cite{t7,t5,t6,t8},  the curvaton model \cite{t9},  theories with
non-canonical kinetic terms
both in single field \cite{t10,t11,t12,t13} and in the multifield case
\cite{miz}, and
in the case in which the cosmological perturbations are induced by vector
perturbations 
of some non-abelian gauge field \cite{dim}. While the most natural value of
the $g_{\rm NL}$
parameter is ${\cal O}(f_{\rm NL}^2)$, there are cases in which $|g_{\rm
NL}|\gg 1$ even if 
 $f_{\rm NL}$ is tiny \cite{enq}. The effects of a cubic correction to the
primordial gravitational 
potential onto the mass function and bias of DM haloes have been recently
analyzed in \cite{ds} 
where the theoretical predictions have been compared to the results extracted
from a series of large $N-$body 
simulations. The limit $-3.5\cdot 10^5<g_{\rm NL}<+8.2\cdot 10^5$ has been
obtained
at 95\% confidence level in the case in which the NG is of the local type. 

The goal of this paper is to present the computation of the DM halo mass
function from the
excursion set theory in 
the presence of a  trispectrum, thus extending our previous computation
of the DM halo mass function when the NG is induced by a
bispectrum \cite{MR3}.
The halo mass function
can be written as
\be\label{dndMdef}
\frac{dn(M)}{dM} = f(\s) \frac{\bar{\rho}}{M^2} 
\frac{d\ln\s^{-1} (M)}{d\ln M}\, ,
\ee
where $n(M)$ is the  number density of dark matter halos of mass $M$,
$\s(M)$ is the variance of the linear density field smoothed on a
scale $R$ corresponding to a mass $M$, and
$\bar{\rho}$ is the average density of the universe. 
Analytical derivations of the halo mass function 
are typically based  on
Press-Schechter (PS) 
theory \cite{PS} and its extension~\cite{PH90,Bond} known as excursion set
theory	(see 
\cite{Zentner} for a recent review).
In  excursion set theory the density
perturbation evolves stochastically 
with the smoothing scale,
and the problem of computing the probability of halo formation is
mapped into the so-called first-passage time problem in the presence
of a barrier. 

The computation of   the effect 
of a primordial  trispectrum  on the mass
function  has been performed   in \cite{MVJ,LoVerde,ds}, and is based  on  
NG extensions of Press-Schechter theory	\cite{PS}.  Performing the
computation
using the two  different  
approximations proposed	in  \cite{MVJ} and \cite{LoVerde},
respectively, 
to evaluate
the impact of a non-vanishing bispectrum on the DM halo mass functions,
one
finds that, to leading order in the large mass limit,
the  predicted halo mass functions are the same for  the two methods, but
differ in the subleading terms, i.e. in the  intermediate mass regime. 
Apart from understanding what is the correct result for the subleading terms, 
there is yet a fundamental reason why we
wish to apply the
excursion set theory to compute the DM mass function with a non-vanishing
trispectrum. 
The derivation of the PS mass function 
in \cite{Bond} requires that the density field $\d$ evolves
with the smoothing scale $R$ (or more precisely with the variance
$S(R)$ of the smoothed density field) in a markovian way. Only under this
assumption one can derive
 the correct factor of two that Press and
Schechter were forced to introduce by hand.
As we have discussed at length in \cite{MR1}, this markovian assumption
is broken by the use of a filter function different from
a sharp filter in momentum space and, of course, it is further
violated by the inclusion of non-Gaussian corrections.  The 
non-markovianity induced by the NG
introduces memory effects which have to be appropriately accounted for
in the excursion set. As we will see, the mass function indeed gets ``memory''
corrections
in the intermediate mass regime, 
which depend on derivatives of the correlators, and therefore cannot be computed with the NG extensions of PS 
theory studied in \cite{MVJ,LoVerde}.

This paper is organized as follows. In Section~\ref{sect:path},
we recall the basic points of the formalism developed in \cite{MR1} for
gaussian fluctuations, and extended in \cite{MR3} to the NG case. 
In Section~\ref{sect:excNG}
we compute the NG corrections with the excursion set method induced by a large
trispectrum, and we
present our results for the halo mass function.
Finally, in Section~\ref{sect:concl} we present our conclusions.

\section{The basic principles of the computation}\label{sect:path}

\subsection{Excursion set theory}

In excursion set theory one
considers the density field $\d$  smoothed over a radius $R$ with a tophat
filter in coordinate space, and studies its stochastic evolution as a function
of the smoothing scale $R$.
As it was found in the classical paper 
\cite{Bond},
when the density $\d(R)$  is smoothed with a sharp filter in momentum
space, and the density fluctuations have gaussian statistics, the smoothed
density field satisfies the equation
\be\label{Langevin1}
\frac{\pa\d(S)}{\pa S} = \eta(S)\, ,
\ee
where $S=\s^2(R)$ is the variance of the linear density field
smoothed on the scale $R$ and computed with a sharp filter in momentum
space, while
$\eta (S)$ is a stochastic variable that satisfies 
\be\label{Langevin2}
\langle \eta(S_1)\eta(S_2)\rangle =\d_D (S_1-S_2)\, ,
\ee
where $\d_D$ denotes the Dirac delta function.
\Eqs{Langevin1} {Langevin2} are   the same as a Langevin equation with a
Dirac-delta  noise $\eta(S)$, with the variance $S$ formally playing the role
of time. 
Let us denote by $\Pi(\d,S)d\d$ the probability density that 
the variable $\d(S)$ reaches a value between $\d$ and $\d+d\d$ by time $S$. 
A textbook  result in statistical physics is that, if a variable $\d(S)$
satisfies a Langevin equation with a Dirac-delta  noise, the  probability
density
$\Pi(\d,S)$ satisfies the Fokker-Planck (FP) equation
\be\label{FPdS}
\frac{\pa\Pi}{\pa S}=\frac{1}{2}\, \frac{\pa^2\Pi}{\pa \d^2}\, .
\ee
The solution of this equation over the whole real axis 
$-\infty<\d<\infty$, with the boundary condition that it vanishes at
$\d=\pm\infty$, is
\be\label{singlegau}
\Pi^0 (\d,S)=\frac{1}{\sqrt{2\pi S}}\, e^{-\d^2/(2S)}\, .
\ee
and is nothing but the distribution function of PS theory. Since large $R$,
i.e. large halo masses, correspond to small values of the variance $S$, in 
ref.~\cite{Bond} it was realized that we are  actually interested in the stochastic
evolution of $\d$ against $S$ only until the ``trajectory" crosses for the first
time  the threshold $\d_c$ for collapse. All the subsequent stochastic
evolution of $\d$ as a function of $S$, which in general results in trajectories
going multiple times above and below the threshold, is irrelevant, since it
corresponds to smaller-scale structures that will be erased and
engulfed by the collapse and virialization of the halo corresponding to the
largest value of $R$, i.e. the smallest value of $S$, for which the threshold
has been crossed.  In other words, trajectories
should be eliminated from further consideration once they have
reached the threshold for the first time. In ref.~\cite{Bond} this is implemented by
imposing the boundary condition
\be\label{bc}
\left.\Pi (\d,S)\right|_{\d=\d_c}=0\, .
\ee
The solution of the FP equation with this boundary condition
is
\be\label{PiChandra}
\Pi (\d,S)=\frac{1}{\sqrt{2\pi S}}\,
\[  e^{-\d^2/(2S)}- e^{-(2\d_c-\d)^2/(2S)} \]
\, ,
\ee
and gives the distribution function of  excursion set theory. 
The first term is the PS result, while the second term in \eq {PiChandra} is 
an ``image" gaussian centered in $\d=2\d_c$.  Integrating this
$\Pi (\d,S)$ over $d\d$ from $-\infty$ to $\d_c$ gives the probability that a
trajectory, at ``time" $S$, has always been below the threshold. Increasing  $S$
this integral decreases because more and more trajectories cross the threshold
for the first time, so the  probability of first
crossing the threshold between ``time'' $S$ and $S+dS$ is given by
${\cal F}(S) dS$, with
\be\label{defcalF}\label{firstcrossT}
{\cal F}(S)  = -\frac{\pa}{\pa S}
\int_{-\infty}^{\d_c}d\d\, \Pi(\d;S)\, .
\ee
With standard manipulations (see e.g. \cite {Zentner} or \cite{MR1}) one then
finds that the function $f(\s)$ which appears in \eq {dndMdef} is given by
\be\label{fcalF}
f(\s)= 2\s^2{\cal F}(\s^2)\, ,
\ee
where we wrote $S=\s^2$. Using \eq {PiChandra} one finds
\be\label{fps}
f(\s) = 
\(\frac{2}{\pi}\)^{1/2}\, 
\frac{\d_c}{\s}\, 
\, e^{-\d_c^2/(2\s^2)}\, ,
\ee
Observe that, when computing the first-crossing rate, the contribution of the
gaussian centered in $\d=0$
and of the image gaussian in \eq {PiChandra} add up, giving the
factor of two that was missed in the original PS theory.

\subsection{Refinements of excursion set theory}

While excursion set theory is quite elegant, and gives a first analytic
understanding of the halo mass function, it suffers of two important set of
problems. First, it is based on the spherical (or ellipsoidal) collapse model,
which is a significant oversimplification of the actual complex dynamics of
halo formation.  We have discussed these limitations in detail in \cite{MR2},
where  we also proposed that some of the physical complications inherent to a
realistic description of halo formation can be included in the excursion set
theory framework, at least at an effective level, by taking into account that
the critical value for collapse is not a fixed constant 
$\d_c$, as in the spherical collapse model, nor a fixed function of the
variance $\s$ of the smoothed density field, as in the ellipsoidal collapse
model, but rather is itself a stochastic variable, whose scatter reflects a
number of complicated aspects 
of the underlying dynamics. The simplest implementation of this idea consists
in 
solving the first-passage time problem in the presence of a  barrier that
performs a random walk, with diffusion coefficient $D_B$, around an average
value given by the constant barrier of the spherical collapse model (more
generally, one should consider a barrier that fluctuates over an average value
given by the ellipsoidal collapse model). In this simple case,
we found  in \cite{MR2} that the exponential factor in the Press-Schechter mass
function changes from $\exp\{-\d_c^2/2\s^2\}$ to $\exp\{-a\d_c^2/2\s^2\}$,
where
$a=1/(1+D_B)$. In this approach all
our
ignorance on the details of halo formation is buried into $D_B$.
The numerical value of $D_B$, and therefore the corresponding value of $a$,
depends among other things also on the details of the algorithm used for
identifying halos (e.g. on the link-length in a Friends-of-Friends algorithm).
Observe that the replacement of
$\exp\{-\d_c^2/2\s^2\}$ with $\exp\{-a\d_c^2/2\s^2\}$ (with $a$ taken 
however as a fitting parameter) 
is just the replacement that was made 
in refs.~\cite{SMT,ST}, in order to fit the results of $N$-body simulations.

The second set of problems of excursion set theory is of a more technical
nature, and is due to the fact that the Langevin equation with Dirac-delta
noise, which is at the basis of the whole construction of ref.~\cite{Bond}, can only
be derived if one works with a sharp filter in momentum space, and if
the fluctuations are gaussian. However, as it
is well known, and as we have discussed at length in ref.~\cite{MR1}, with such a
filter it is not possible to associate a halo mass to the smoothing scale $R$.
A unambigous relation between $M$ and $R$ is rather obtained with a sharp
filter in {\em coordinate} space, in which case one simply has $M=(4/3)\pi
R^3\rho$, where $\rho$ is the density.
When one uses a sharp filter in coordinate space, the evolution of the density
with the smoothing scale however becomes non-markovian, and therefore the
problem becomes technically much more difficult. In particular, the distribution
function $\Pi(\d,S)$ no longer satisfies a local differential equation such as
the FP equation. The issue is particularly relevant when one wants to include
non-Gaussianities in the formalism, since 
again the inclusion of non-Gaussianities renders the dynamics non-markovian. 

In refs.~\cite{MR1,MR3}  we have developed a formalism that allows us to
generalize excursion set theory to the case of a non-markovian dynamics, either
generated by the filter function or by primordial non-Gaussianities. The basic
idea is the following. Rather than trying to derive a simple, local,
differential equation for $\Pi(\d,S)$ (which, as we have shown in ref.~\cite{MR1},
is impossible; in the non-markovian case $\Pi(\d,S)$  rather satisfies a very
complicated equation which is non-local with respect to ``time" $S$), we
construct the probability distribution $\Pi(\d,S)$ directly by summing over all
paths that never exceeded the threshold $\d_c$, i.e. 
by writing $\Pi(\d,S)$ as a
path integral with boundaries.	To obtain such a representation, 
we consider an ensemble of
trajectories all starting at $S_0=0$ from an initial position
$\d(0) =\d_0$ 
and we follow them for a ``time'' $S$.	
We discretize the interval $[0,S]$ in steps
$\D S=\eps$, so $S_k=k\eps$ with $k=1,\ldots n$, and $S_n\equiv S$. 
A trajectory is  then defined by
the collection of values $\{\d_1,\ldots ,\d_n\}$, such that $\d(S_k)=\d_k$.
The probability density in the space of  trajectories is 
\be\label{defW}
W(\d_0;\d_1,\ldots ,\d_n;S_n)\equiv \langle
\d_D (\d(S_1)-\d_1)\ldots \d_D (\d(S_n)-\d_n)\rangle\, ,
\ee
where  $\d_D$
denotes the Dirac delta. Then the probability of arriving in $\d_n$ in a
``time'' $S_n$, starting from an initial value $\d_0$, without ever
going above the threshold, is\footnote{In \eqs
{singlegau} {PiChandra} we implicitly assumed $\d_0=0$. In the following
however it will be necessary to keep track also of the initial position $\d_0$.}
\be\label{defPi}
\Pi_{\eps} (\d_0;\d_n;S_n)
 \equiv\int_{-\infty}^{\d_c} d\d_1\ldots \int_{-\infty}^{\d_c}d\d_{n-1}\, 
W(\d_0;\d_1,\ldots ,\d_{n-1},\d_n;S_n).
\ee
The label $\eps$ in $\Pi_{\eps}$ reminds us that this quantity is defined with
a finite spacing $\eps$, and we are finally interested in the continuum limit
$\eps\ra 0$.
As we discussed in \cite{MR1,MR3}, $W(\d_0;\d_1,\ldots ,\d_{n-1},\d_n;S_n)$ can
be expressed in terms of the connected correlators of the theory,
\bees\label{WnNG}
&&W(\d_0;\d_1,\ldots ,\d_n;S_n)=\Dl
\\
&&\exp\left\{ i\sum_{i=1}^n\lambda_i\d_i+\sum_{p=2}^{\infty}
\frac{(-i)^p}{p!}\,
 \sum_{i_1=1}^n\ldots \sum_{i_p=1}^n
\lambda_{i_1}\ldots\lambda_{i_p}\,
\langle \d_{i_1}\ldots\d_{i_p}\rangle_c
\right\}\, ,\nn
\ees
where
\be
\Dl \equiv
\inT\frac{d\lambda_1}{2\pi}\ldots\frac{d\lambda_n}{2\pi}\, ,
\ee
$\d_i=\d(S_i)$,
and $\langle \d_{1}\ldots\d_{n}\rangle_c$ denotes the connected $n$-point
correlator. So
\bees\label{Piexplicit}
&&\Pi_{\eps}(\d_0;\d_n;S_n)=\int_{-\infty}^{\d_c} d\d_1\ldots d\d_{n-1}\,
\Dl\\
&&\exp\left\{ i\sum_{i_1=1}^n\lambda_i\d_i+\sum_{p=2}^{\infty}
\frac{(-i)^p}{p!}\, 
\sum_{i=1}^n\ldots \sum_{i_p=1}^n
\lambda_{i_1}\ldots\lambda_{i_p}\,
\langle \d_{i_1}\ldots\d_{i_p}\rangle_c
\right\}\, .\nn
\ees
When $\d(S)$ satisfies \eqs {Langevin1} {Langevin2} (which is the case
for sharp filter in momentum space) the two-point function can be easily
computed, and is given by
\be\label{twopoint}
\langle\delta(S_i)\delta(S_j)\rangle = {\rm min}(S_i,S_j)\, .
\ee
If furthermore we consider gaussian fluctuations,  all $n$-point connected
correlators with $n\geq 3$ vanish, and 
the probability density $W$ can be computed explicitly,
\be\label{W}
W^{\rm gm}(\d_0;\d_1,\ldots ,\d_n;S_n)=\frac{1}{(2\pi\eps)^{n/2}}\, 
\exp\left\{-\frac{1}{2\eps}\,
\sum_{i=0}^{n-1}  (\d_{i+1}-\d_i)^2\right\},
\ee
where the superscript ``gm'' (gaussian-markovian) reminds that this value of
$W$ is computed for gaussian fluctuations, whose dynamics with respect to the
smoothing scale is markovian.
Using this result, in \cite{MR1} we have shown that, in the continuum
limit, the distribution function $\Pi_{\eps=0}(\d;S)$, computed
with a sharp filter in
momentum space, satisfies a Fokker-Planck equation with the boundary
condition $\Pi_{\eps=0}(\d_c,S)=0$, and we have therefore recovered, from our
path integral approach, the distribution function of excursion set
theory,  \eq {PiChandra}.

When the consider a different filter, \eq {twopoint} is replaced by
\be\label{twopoint2}
\langle\delta(S_i)\delta(S_j)\rangle = {\rm min}(S_i,S_j)+\Delta (S_i,S_j)\, ,
\ee
where $\Delta (S_i,S_j)$ describes the deviations from a markovian dynamics.
For instance, for a sharp filter in coordinate space, which is the most
interesting case, 
the function $\Delta (S_i,S_j)$ is very well approximated by 
\be\label{approxDelta}
\D(S_i,S_j)\simeq \kappa\, \frac{S_i(S_j-S_i)}{S_j}\, ,
\ee
(for $S_i\leq S_j$, and is symmetric under exchange of $S_i$ and $S_j$),
with $\kappa\simeq 0.45$. The non-markovian corrections can then be computed
expanding perturbatively in $\kappa$. The computation, which is quite
non-trivial from a technical point of view, has been discussed in great detail
in \cite{MR1}. Let us summarize here the crucial points. First of all,
expanding to first order in $\D_{ij}$
and using
$\lambda_ie^{i\sum_k\lambda_k \d_k}=-i\pa_ie^{i\sum_k\lambda_k\d_k}$, 
where $\pa_i=\pa/\pa\d_i$, the first-order correction to
$\Pi_{\eps}$ is
\bees
&&\Pi^{{\D}1}_{\eps}(\delta_0;\delta_n;S_n) \equiv
\int_{-\infty}^{\delta_c} 
d\delta_1\ldots d\delta_{n-1}\,\frac{1}{2}\sum_{i,j=1}^n \D_{ij}\pa_i\pa_j\nn\\
&&\times
\int{\cal D}\lambda\, 
\exp\left\{
i\sum_{i=1}^n\lambda_i\delta_i -\frac{1}{2}\,\sum_{i,j=1}^n {\rm min}(S_i,S_j)
\lambda_i\lambda_j\right\}\label{Delta3}\\
&&=\frac{1}{2}\sum_{i,j=1}^n \D_{ij}\int_{-\infty}^{\delta_c} 
d\delta_1\ldots d\delta_{n-1}\,\pa_i\pa_j
W^{\rm gm}(\delta_0;\delta_1,\ldots ,\delta_n;S_n)
\, ,\nn
\ees
where we use the notation $\D_{ij}=\D(S_i,S_j)$.
One then observes that the derivatives  $\pa_i$ run over
$i=1,\ldots ,n$, while we integrate only over $d\delta_1\ldots d\delta_{n-1}$.
Therefore, derivatives $\pa_i$ with $i=n$ can be simply carried outside the
integrals. Derivatives $\pa_i$ with $i=1,\ldots, n-1$ 
are dealt as follows. Consider for instance the terms with $i<n$ and
$j=n$ (together with $j<n$ and $i=n$, which gives a factor of
two). These are given by
\be\label{expr}
\sum_{i=1}^{n-1}\D_{in}\pa_n\int_{-\infty}^{\delta_c} 
d\delta_1\ldots d\delta_{n-1}\,\pa_iW^{\rm gm}(\delta_0;\delta_1,\ldots
,\delta_n;S_n)\, ,\nn
\ee
To compute this expression we  integrate $\pa_i$ by parts,
\bees
&&\int_{-\infty}^{\delta_c} 
d\delta_1\ldots d\delta_{n-1}\,\pa_iW^{\rm gm}(\delta_0;\delta_1,\ldots
,\delta_n;S_n)\label{byparts}\\
&&=\int_{-\infty}^{\delta_c} d\delta_1\ldots \widehat{d\d}_i\ldots 
d\delta_{n-1}
W(\delta_0;\delta_1,\ldots, \delta_i=\delta_c,\ldots
,\delta_{n-1},\delta_n;S_n)\, ,\nn
\ees
where
the notation $ \widehat{d\d}_i$ means that we must omit $d\delta_i$ from
the list of integration variables. 
We next observe that $W^{\rm gm}$ satisfies
\bees\label{facto}
&&W^{\rm gm}(\delta_0;\delta_1,\ldots, \delta_{i}= \delta_c, 
\ldots ,\delta_n; S_n)\\
&&= 
W^{\rm gm}(\delta_0;\delta_1,\ldots, \delta_{i-1}, \delta_c;S_i)
W^{\rm gm}(\delta_c; \delta_{i+1}, \ldots ,\delta_n; S_n-S_i)\, ,\nn
\ees
as can be verified directly from its explicit expression (\ref{W}). Then
\bees\label{WPiPi}
&&\int_{-\infty}^{\delta_c}d\delta_1\ldots d\delta_{i-1}
\int_{-\infty}^{\delta_c}  d\delta_{i+1}\ldots	d\delta_{n-1}\nn\\
&&\times W^{\rm gm}(\delta_0;\delta_1,\ldots, \delta_{i-1}, \delta_c;S_i)
W^{\rm gm}(\delta_c; \delta_{i+1}, \ldots ,\delta_n; S_n-S_i)\nn\\
&&= \Pi^{\rm gm}_{\eps}(\delta_0;\delta_c;S_i)
\Pi^{\rm gm}_{\eps}(\delta_c;\delta_n;S_n-S_i)\, ,
\ees
and to compute the expression given in \eq{expr} we must compute
\be\label{Pimem1}
\sum_{i=1}^{n-1}\D_{in}
 \Pi^{\rm gm}_{\eps}(\delta_0;\delta_c;S_i)
\Pi^{\rm gm}_{\eps}(\delta_c;\delta_n;S_n-S_i) .
\ee
To proceed further, we need to know $\Pi^{\rm
gm}_{\eps}(\delta_0;\delta_c;S_i)$. By definition, for $\eps=0$ this quantity
vanishes, since its second argument is equal to the the threshold value $\d_c$,
compare with \eq{bc}. However, in the continuum limit the sum over $i$ becomes
$1/\eps$ times an integral over an intermediate time variable $S_i$,
\be\label{sumint}
\sum_{i=1}^{n-1}\ra \frac{1}{\eps}\int_o^{S_n} dS_i\, ,
\ee
so we need to know how $\Pi^{\rm gm}_{\eps}(\delta_0;\delta_c;S_i)$ approaches
zero when $\eps\ra 0$.	In \cite{MR1} we proved that it vanishes as
$\sqrt{\eps}$, and that
\be\label{Pigammafinal}
\Pi^{\rm gm}_{\eps} (\delta_0;\delta_c;S)=
\sqrt{\eps}\, \frac{1}{\sqrt{\pi}}\, 
\frac{\delta_c-\delta_0}{S^{3/2}} e^{-(\delta_c-\delta_0)^2/(2S)} +{\cal
O}(\eps)\, .
\ee
Similarly, for $\delta_n<\delta_c$,
\be\label{Pigammafinalbis}
\Pi^{\rm gm}_{\eps} (\delta_c;\delta_n;S)=
\sqrt{\eps}\, \frac{1}{\sqrt{\pi}}\, 
\frac{\delta_c-\delta_n}{S^{3/2}} e^{-(\delta_c-\delta_n)^2/(2S)} +{\cal
O}(\eps)\, .
\ee
Therefore the two factor $\sqrt{\eps}$ from 
\eqs{Pigammafinal}{Pigammafinalbis} produce just an 
overall factor of $\eps$ that
compensate the factor $1/\eps$ in \eq {sumint}, and we are left with a finite
integral over $dS_i$.  Terms with two or more derivative, e.g. $\pa_i\pa_j$, or
$\pa_i,\pa_j\pa_k$ acting on $W$, with all indices $i,j,k$ maller than $n$, can
be computed similarly, and  have been discussed in detail in
\cite{MR1}.

\section{Contribution of the trispectrum to the halo 
mass function}\label{sect:excNG}

The effect of non-Gaussianities can be computed similarly,
expanding perturbatively 
\eq {Piexplicit} in terms of the higher-order correlators. In
\cite{MR3} we have examined  the three-point correlator, i.e. 
the bispectrum. Here we compute the effect of the trispectrum.

If in \eq{Piexplicit}
we only retain the four-point correlator, and we use the tophat filter in
coordinate space, we have
\bees
&&\Pi_{\eps}(\d_0;\d_n;S_n) =
\int_{-\infty}^{\d_c} 
d\d_1\ldots d\d_{n-1}\,\Dl\label{Pi3}\nonumber\\
&&\times
\exp\left\{i\lambda_i\d_i -\frac{1}{2}
[{\rm min}(S_i,S_j) + \D_{ij}]
\lambda_i\lambda_j
+\frac{(-i)^4}{24}\langle\d_i\d_j\d_k\d_l
\rangle\lambda_i\lambda_j\lambda_k\lambda_l
\right\}.
\ees
Expanding to first order, $\D_{ij}$ and 
$\langle\d_i\d_j\d_k\d_l\rangle$ do not mix, so we must compute
\be
\Pi^{(4)}_{\eps}(\d_0;\d_n;S_n)\equiv 
\frac{1}{24}\sum_{i,j,k,l=1}^n
\langle\d_i\d_j\d_k\d_l\rangle\int_{-\infty}^{\d_c} 
d\d_1\ldots d\d_{n-1}\pa_i\pa_j\pa_k\pa_l
W^{\rm gm}\, ,\label{Pi4full}
\ee
where the superscript $(4)$ in $\Pi^{(4)}_{\eps}$
refers to the fact that this is the
contribution linear in the four-point correlator. 

In principle this expression can be computed with the same technique
discussed above, separating the various contributions to the sum
according to whether an index is equal or smaller than $n$. In this
way, however, the computations faces some technical
difficulties.\footnote{In particular, when we have terms with three or
  more derivatives, we need to generalize \eq{Pigammafinalbis},
  including terms up to ${\cal O}(\eps^{3/2})$, which is quite 
non-trivial.}
Fortunately, the problem simplifies considerably in the limit of large
halo masses, which is just the physically interesting limit. Large
masses mean small values of $S_n$. In \eq{Pi4full} the arguments
$S_i, S_j,S_k$ and
$S_l$ in the correlator
$\langle\d_i\d_j\d_k\d_l\rangle \equiv
\langle\d(S_i)\d(S_j)\d(S_k)\d(S_l)\rangle$
range over the interval $[0,S_n]$ and, if $S_n$ goes to zero, we
can expand the correlator in a multiple Taylor series
around the point $S_i=S_j=S_k=S_l=S_n$.
We introduce the notation
\be
G_4^{(p,q,r,s)}(S_n)\equiv
\[\frac{d^p}{dS_i^p}\frac{d^q}{dS_j^q}\frac{d^r}{dS_k^r}\frac{d^s}{dS_l^s}
\langle\d(S_i)\d(S_j)\d(S_k)\d(S_l)\rangle\]_{S_i=S_j=S_k=S_l=S_n}
\, .
\ee
Then
\bees\label{pqrs}
\langle\d(S_i)\d(S_j)\d(S_k)\d(S_l)\rangle
&=&
\sum_{p,q,r,s=0}^{\infty} \frac{(-1)^{p+q+r+s}}{p!q!r!s!}\\
&&\hspace*{5mm}\times
(S_n-S_i)^{p}(S_n-S_j)^{q}
(S_n-S_k)^{r}(S_n-S_l)^{s}G_3^{(p,q,r,s)}(S_n)\, .\nn
\ees
Terms
with more and more derivatives give contributions to the function
$f(\s)$, defined in \eq{dndMdef}, that are subleading in the limit of
small $\s$, i.e. for $\s/\d_c\ll 1$,
and the leading contribution to the halo mass function 
is given by the term in \eq{pqrs} with $p=q=r=s=0$. At
next-to-leading order we must also include
the contribution of the terms in \eq{pqrs}
with $p+q+r+s=1$, i.e. the four terms
$(p=1,q=0,r=0,s=0)$, $(p=0,q=1,r=0,s=0)$,  $(p=0,q=0,r=1,s=0)$ and
$(p=0,q=0,r=0,s=1)$; at
next-to-next-to-leading order
we must include the
contribution of the terms in \eq{pqrs}
with $p+q+r=2$, and so on.
For the
purpose of organizing the expansion in leading term, subleading
terms, etc., we can
reasonably expect that, for small $S_n$
\be\label{ordering}
G_4^{(p,q,r,s)}(S_n)\sim
S_n^{-(p+q+r+s)}\langle\d^{4}(S_n)\rangle\, ,
\ee
i.e. each derivative $\pa/\pa S_i$,  when evaluated in $S_i=S_n$, gives a
factor of order $1/S_n$.
This ordering  will be assumed when we present our
final result for the halo mass function below. However, our formalism
allows us  to compute each contribution 
separately, so our results below can be easily generalized in order to
cope with a different hierarchy between the various $G_4^{(p,q,r,s)}(S_n)$.

The leading term in $\Pi^{(4)}$ is
\be\label{Pi4appr}
\Pi^{(4,{\rm L})}_{\eps}(\d_0;\d_n;S_n)= 
\frac{\langle\d_n^4\rangle}{24}\, 
\sum_{i,j,k,l=1}^n
\int_{-\infty}^{\d_c} 
d\d_1\ldots d\d_{n-1}\pa_i\pa_j\pa_k\pa_l
W^{\rm gm}\, ,\nn
\ee
where the superscript ``${\rm L}$'' stands for ``leading''.
Since in the end we are interested in  the integral over $d\d_n$ of
$\Pi_{\eps}(\d_0;\d_n;S_n)$, see \eq{defcalF}, we can write directly
\be
\int_{-\infty}^{\d_c}d\d_n\, \Pi^{(4,{\rm L})}_{\eps=0}(0;\d_n;S_n)
=\frac{\langle\d^4(S_n)\rangle}{24}\,
\sum_{i,j,k,l=1}^n
\int_{-\infty}^{\d_c} 
d\d_1\ldots d\d_{n}\pa_i\pa_j\pa_k\pa_l
W^{\rm gm}\, .\label{eq34}
\ee
This expression
can  be computed very easily  by making use of identities that we 
proved in \cite{MR1,MR3}. Namely, we consider the derivative 
of $\Pi^{\rm gm}_{\eps}$ with
respect to the threshold $\d_c$ (which, when we use the notation
$\Pi^{\rm gm}_{\eps}(\d_0;\d_n;S_n)$,
is not written explicitly in the list of
variable on which  $\Pi^{\rm gm}_{\eps}$ depends, but of course
enters as upper integration limit in
\eq{defPi}). Then one can show that
\be\label{pa1xc}
\sum_{i=1}^{n}
\int_{-\infty}^{\d_c} d\d_1\ldots d\d_{n}\, 
\pa_iW^{\rm gm}=
\frac{\pa}{\pa \d_c}
\int_{-\infty}^{\d_c} d\d_n\,\Pi^{\rm gm}_{\eps} 
\, ,
\ee
\be\label{pa2xc}
\sum_{i,j=1}^{n}
\int_{-\infty}^{\d_c} d\d_1\ldots d\d_{n}\, 
\pa_i\pa_jW^{\rm gm}=
\frac{\pa^2}{\pa \d_c^2}
\int_{-\infty}^{\d_c} d\d_n\,\Pi^{\rm gm}_{\eps} 
\, ,
\ee
and similarly for all higher-order derivatives, so in particular
\be\label{pa4xc}
\sum_{i,j,k,l=1}^{n}
\int_{-\infty}^{\d_c} d\d_1\ldots d\d_{n}\, 
\pa_i\pa_j\pa_k\pa_lW^{\rm gm}=
\frac{\pa^4}{\pa \d_c^4}
\int_{-\infty}^{\d_c} d\d_n\,\Pi^{\rm gm}_{\eps} 
\, .
\ee
Therefore, in the
continuum limit, the right-hand side of \eq{eq34} 
is computed very simply, just by inserting in \eq{pa4xc} the value 
of $\Pi^{\rm gm}_{\eps}$ for $\eps\ra 0$,
\be\label{Pigaueps0}
\Pi^{\rm gm}_{\eps=0} (\d_0=0;\d_n;S_n)
= \frac{1}{\sqrt{2\pi S_n}}\,
\[e^{-\d_n^2/(2S_n)}-
e^{-(2\d_c-\d_n)^2/(2S_n)}\]\, ,
\ee
and therefore, in the continuum limit,
\be\label{Pi4finite}
\int_{-\infty}^{\d_c}d\d_n\, \Pi^{(4,{\rm L})}_{\eps=0}(0;\d_n;S_n)
=\frac{\langle\d_n^4\rangle}{12\sqrt{2\pi}\, S_n^{5/2}}
\delta_c\( 3-\frac{\d_c^2}{S_n}\)
\, e^{-\d_c^2/(2S_n)}\, .
\ee
We now insert this result into \eqs{defcalF}{fcalF}
and we express the result in terms of the normalized kurtosis
\be\label{defkurt}
{\cal S}_4(\s)\equiv \frac{1}{S^3}\langle\d^4(S)\rangle\, .
\ee
Putting the contribution of $\Pi^{(4,{\rm L})}$ 
together with the gaussian contribution, and writing $S=\s^2$,
we find
\be\label{fsR1ds}
f(\s)= \(\frac{2}{\pi}\)^{1/2}\, 
\frac{\d_c}{\s}\, \, e^{-\d_c^2/(2\s^2)} \left[ 1+\frac{\s^2{\cal S}_4(\s)}{24}
\(\frac{\d_c^4}{\s^4}-\frac{4\d_c^2}{\s^2}-3\)
+\frac{1}{24}\s^2\frac{d{\cal S}_4}{d\ln\s}\,
\(\frac{\d_c^2}{\s^2}-3\)\right]\, .
\ee
Let us emphasize that    the  variance $\s^2$ is 
 to  be computed applying the linear transfer function to  the primordial gravitational
potential (\ref{phi}) containing the extra $g_{\rm NL}$-piece. The result 
given in \eq{fsR1ds}
agrees with the one
obtained in \cite{LoVerde}
by performing the  Edgeworth expansion of a
non-Gaussian generalization of Press-Schechter theory. 
However, just as we have discussed in \cite{MR3} for the case of the
bispectrum, \eq{fsR1ds} cannot be  taken as the full result
beyond leading order. If we want to compute consistently to NL order,
we need to include the terms with $p+q+r+s=1$ in
\eq{pqrs}, which is given by
\begin{eqnarray}\label{Pi4apprcNL}
\int_{-\infty}^{\d_c}  d\d_n\, \Pi^{(4,{\rm NL})}_{\eps}(\d_0;\d_n;S_n)
&=&-\frac{4}{24}\,G_4^{(1,0,0,0)}(S_n)\sum_{i=1}^n (S_n-S_i)\nn\\
 &&\times\sum_{j,k,l=1}^n
\int_{-\infty}^{\d_c} 
d\d_1\ldots d\d_{n-1}d\d_n\pa_i\pa_j\pa_k\pa_l
W^{\rm gm}\, ,
\end{eqnarray}
where the superscript ``NL'' in $\Pi^{(4,{\rm NL})}_{\eps}$
stands for next-to-leading,
and we used the fact that the four terms
$(p=1,q=0,r=0,s=0), \ldots ,
(p=0,q=0,r=0,s=1)$ give the same
contribution. 
We now use the same trick as before to eliminate
$\sum_{j,k,l=1}^n\pa_j\pa_k\pa_k$ in favor of $\pa^3/\pa \d_c^3$, 
\be\label{Pi4apprc}
\int_{-\infty}^{\d_c}  d\d_n\Pi^{(4,{\rm NL})}_{\eps}(\d_0;\d_n;S_n)=
-\frac{4}{24}\,
G_4^{(1,0,0,0)}(S_n)
\sum_{i=1}^n (S_n-S_i)\frac{\pa^3}{\pa \d_c^3}
\int_{-\infty}^{\d_c} 
d\d_1\ldots d\d_{n-1}d\d_n\pa_i
W^{\rm gm}\, .
\ee
The remaining path integral can be computed using the technique discussed in
\eqst{byparts}{Pigammafinalbis}, and we get
\bees\label{Pi4apprcinter}
&&\int_{-\infty}^{\d_c}  d\d_n\, \Pi^{(4,{\rm NL})}_{\eps}(\d_0;\d_n;S_n)
=-\frac{4}{24\pi}\, G_4^{(1,0,0,0)}(S_n)\int_0^{S_n}dS_i\, 
\frac{1}{S_i^{3/2} (S_n-S_i)^{1/2}}\nn\\
&&\times\frac{\pa^3}{\pa \d_c^3}\, 
\[ \d_c e^{-\d_c^2/(2S_i)}\int_{-\infty}^{\d_c}d\d_n\,
(\d_c-\d_n) \exp\left\{-\frac{(\d_c-\d_n)^2}{2(S_n-S_i)}\right\} \]
\, .
\ees
Observe that this expression involves an integral over all values of an intermediate 
``time" variable $S_i$ which ranges over $0\leq S_i\leq S_n$. As we have discussed
in detail in \cite{MR1,MR3} these terms are ``memory" 
terms that depend on the whole past history of the trajectory, 
and reflect the non-markovian nature of the stochastic process.

The integral over $d\d_n$ is easily performed writing
\be
(\d_c-\d_n) \exp\left\{-\frac{(\d_c-\d_n)^2}{2(S_n-S_i)}\right\}
=(S_n-S_i)\pa_n\exp\left\{-\frac{(\d_c-\d_n)^2}{2(S_n-S_i)}\right\}
\, ,
\ee
so it just gives $(S_n-S_i)$.
Carrying out the third derivative
with respect to $\d_c$ and the remaining elementary integral over
$dS_i$	we  get
\be\label{Pi4NL}
\int_{-\infty}^{\d_c}  d\d_n\, \Pi^{(4,{\rm NL})}_{\eps}(\d_0;\d_n;S_n)
=\frac{\d_c}{3\sqrt{2\pi}}\, \frac{G_4^{(1,0,0,0)}(S_n)}{S_n^{3/2}}
\, e^{-\d_c^2/(2S_n)}\, .
\ee
We now define
\be
{\cal U}_4(\s)\equiv \frac{4 G_4^{(1,0,0,0)}(S)}{S^2}\, ,
\ee
where as usual $S=\s^2$.
When the ordering given in \eq{ordering} holds, ${\cal U}_4(\s)$ is of the same
order
as the normalized kurtosis ${\cal S}_4(\s)$.
Computing the contribution to $f(\s)$ from
\eq{Pi4NL} and we finally find
\bees\label{fsR6}
&&f(\s)= \(\frac{2}{\pi}\)^{1/2}\, 
\frac{\d_c}{\s}\, \, e^{-\d_c^2/(2\s^2)}
 \left[ 1+\frac{\s^2{\cal S}_4(\s)}{24}
\(\frac{\d_c^4}{\s^4}-\frac{4\d_c^2}{\s^2}-3\)
+\frac{1}{24}\s^2\frac{d{\cal S}_4}{d\ln\s}\,
\(\frac{\d_c^2}{\s^2}-3\)\right.\nn\\ 
&&\phantom{\(\frac{2}{\pi}\)^{1/2}\, 
\frac{\d_c}{\s}\, \, e^{-\d_c^2/(2\s^2)}}
\hspace*{15mm}\left. -\frac{\s^2{\cal U}_4(\s)}{24}\left(\frac{\d_c^2}{\s^2}+1\right)-
\frac{1}{24}\s^2\frac{d{\cal U}_4}{d\ln\s}\right]\,.
\ees
The above result only holds up to NL order. If one wants to use it up
to NNL order, the terms in square bracket where $\s^2{\cal S}_4(\s)$,
$\s^2d{\cal S}_4/d\ln\s$,  $\s^2{\cal U}_4(\s)$,
and $\s^2d{\cal U}_4/d\ln\s$ are multiplied by factors of ${\cal O}(1)$
must be supplemented by the computation of the terms with 
$p+q+r+s=2$ in \eq{pqrs}, which will give a contribution 
analogous to the term ${\cal V}_3(\s)$ computed in \cite{MR3}. However,
with present numerical accuracy,
NNL terms are not yet relevant for the comparison with $N$-body
simulations with non-Gaussian initial conditions.

Our result may be refined  in several ways.
Until now we have worked with a barrier with a fixed height $\d_c$
and we neglected the corrections due to the filter.
We can now include the modifications due to the fact that the height of
the barrier may be thought to diffuse stochastically, as discussed in
\cite{MR2}, and
also the corrections due to the filter.
To compute the non-Gaussian term proportional to 
the four-point correlator with the diffusing barrier we recall, from
\cite{MR2}, that the first-passage time problem of a particle obeying a
diffusion equation with diffusion coefficient $D=1$, in the presence of a
barrier that moves stochastically with diffusion coefficient $D_B$,
can be mapped into the first-passage time problem of a particle with
effective diffusion coefficient $(1+D_B)$, and fixed barrier. This can be
reabsorbed into a rescaling of the ``time'' variable $S\ra (1+D_B)S=S/a$,
and therefore $\s\ra \s/\sqrt{a}$. At the same time 
the four-point correlator must be rescaled according to
$\langle\d_n^4\rangle\ra a^{-2}\langle\d_n^4\rangle$ since,
dimensionally, $\langle\d_n^4\rangle$ is the same as $S^{2}$, which means
that
${\cal S}_4\ra a{\cal S}_4$. 
As a final ingredient, we must add the effect of the tophat filter
function in coordinate space. 
For a tophat filter in coordinate space, we have found \cite{MR1} that the
two-point 
correlator is given by \eqs{twopoint2}{approxDelta}.
Including the 
non-markovianity induced by the tophat smoothing function in real
space, using the  computations already performed in \cite{MR1,MR3}, we end
up with 
\bees\label{fsR5} 
&&f(\s)= (1-\tilde{\kappa})\(\frac{2}{\pi}\)^{1/2}\, 
\frac{a^{1/2}\d_c}{\s}\, \, e^{-a\d_c^2/(2\s^2)}\nn\\
&&\times \left[ 1+\frac{\s^2{\cal S}_4(\s)}{24}
\(\frac{a^2\d_c^4}{\s^4}-\frac{4a\d_c^2}{\s^2}-3\)
+\frac{1}{24}\s^2\frac{d{\cal S}_4}{d\ln\s}\,
\(\frac{a\d_c^2}{\s^2}-3\)\right.\nn\\
&&\left.-\frac{\s^2{\cal U}_4(\s)}{24}\left(\frac{a\d_c^2}{\s^2}+1\right)-
\frac{1}{24}\s^2\frac{d{\cal U}_4}{d\ln\s}\right]
+\frac{\tilde{\kappa}}{\sqrt{2\pi}}\,
\frac{a^{1/2}\d_c}{\s}\,
 \G\(0,\frac{a \d_c^2}{2\s^2}\)\, , 
\ees
where $\tilde{\kappa}=\kappa/(1+D_B)$.
This is our final result. 
More generally, also the term proportional to the incomplete Gamma function
could get non-Gaussian corrections, which in principle can be computed
evaluating perturbatively a ``mixed'' term proportional to 
$\D_{ij}\langle\d_k\d_l\d_m\d_n\rangle\pa_i\pa_j\pa_k\pa_l\pa_m\pa_n$.
However we saw in \cite{MR1} that 
in the large mass limit, where the
non-Gaussianities are important, the term proportional to the incomplete
Gamma function is subleading, so we will neglect the non-Gaussian
corrections to this subleading term.

\section{Conclusions}\label{sect:concl}
In this paper we have computed the DM halo mass function as predicted within
the excursion set theory when
a  NG is present under the form of a  trispectrum. We  thus have
extended our previous results presented in ref.~\cite{MR3},
where a similar computation was performed in the presence of a  NG
bispectrum. 
Our computation  accounts for the non-markovianity of the random walk
of the smoothed density
contrast, which inevitably arise when  deviations from gaussianity are present.
While our result coincides at the leading order
${\cal O}(\d_c^4/\s^4)$ 
with that obtained in \cite{LoVerde} through  PS theory, it is different at the
order ${\cal O}(\d_c^2/\s^2)$. This is 
due to the memory effects induced by the non-markovian excursion set which are
not present in the PS approach. 
Our final expression  (\ref{fsR5}) takes into account as well the non-markovian
effects due to the choice of the
tophat filter in real space and  the proper exponetial decay at large DM
masses.   
\vspace{5mm}

\noindent
The work
of MM is supported by the Fond National Suisse. 
The work of AR is supported by 
the European Community's Research Training Networks 
under contract MRTN-CT-2006-035505.

\bibliographystyle{JHEP}

\end{document}